\documentclass
[prl,twocolumn,superscriptaddress,floats,showpacs]{revtex4}%
\usepackage{amsfonts}
\usepackage{url}
\usepackage{natbib}
\usepackage{float}
\usepackage{amsmath}
\usepackage{amssymb}
\usepackage{graphicx}%

\begin{document}
\preprint{cond-mat/}
\title[Short title for running header]{Zener double exchange from local valence fluctuations in magnetite}
\author{R. J. McQueeney}
\affiliation{Department of Physics \& Astronomy, Iowa State University, Ames, Iowa 50011 USA}
\affiliation{Ames Laboratory, Ames, Iowa 50011 USA}
\author{M. Yethiraj}
\altaffiliation{current address: Bragg Institute, ANSTO, Lucas Heights, NSW 2234 Australia}

\affiliation{Center for Neutron Scattering, Oak Ridge National Laboratory, Oak Ridge, TN
37831 USA}
\author{S. Chang}
\affiliation{Ames Laboratory, Ames, Iowa 50011 USA}
\author{W. Montfrooij}
\affiliation{Department of Physics and Missouri Research Reactor, University of Missouri,
Columbia, MO 65211 USA}
\author{T. G. Perring}
\affiliation{ISIS Facility, Rutherford-Appleton Laboratory, Clarendon, UK}
\author{J. M. Honig}
\affiliation{Department of Chemistry, Purdue University, West Lafayette, IN 47907 USA}
\author{P. Metcalf}
\affiliation{Department of Chemistry, Purdue University, West Lafayette, IN 47907 USA}


\begin{abstract}
Magnetite (Fe$_{3}$O$_{4}$) is a mixed valent system where electronic
conductivity occurs on the B-site (octahedral) iron sublattice of the spinel
structure. Below \thinspace$T_{V}=122$ K, a metal-insulator transition occurs
which is argued to arise from the charge ordering of 2+ and 3+ iron valences
on the B-sites (Verwey transition). Inelastic neutron scattering measurements
show that optical spin waves propagating on the B-site sublattice ($\sim$80
meV) are shifted upwards in energy above $T_{V}$ due to the occurrence of B-B
ferromagnetic double exchange in the mixed valent metallic phase. The double
exchange interaction affects only spin waves of $\Delta_{5}$ symmetry, not all
modes, indicating that valence fluctuations are slow and the double exchange is
constrained by electron correlations above $T_{V}$.
\end{abstract}
\date{July 15, 2007}
\maketitle
Magnetite (Fe$_{3}$O$_{4}$) is the prototypical example of a metal-insulator
transition with a charge-ordered (CO) insulating ground state (Verwey
transition). Since its discovery nearly 70 years ago,\cite{verwey1939} the
driving forces behind the Verwey transition are still not completely
understood.\cite{walz2002,garcia2004} Magnetite has a cubic inverse spinel
crystal structure containing two different symmetry iron sites; the A-site
resides in tetrahedrally coordinated oxygen interstices and has stable valence
(3d$^{5}$, Fe$^{3+}$), the two B-sites have octahedral coordination and a
fractional average valence of 2.5+. The ferrimagnetic structure consists of
ferromagnetic A- and B-sublattices aligned antiparallel to each other ($T_{C}$
= 858 K). Below $T_{V}$ = 122 K, magnetite undergoes a metal-insulator
transition resulting in a decrease of the conductivity by two
orders-of-magnitude. The model that has persisted over time is that extra
electrons forming Fe$^{2+}$ ions (3d$^{6}$) hop to neighboring Fe$^{3+}$ sites
on the tetrahedral B-sublattice network and give rise to electrical
conductivity. Anderson argued that short-range ordering of Fe$^{2+}$ and
Fe$^{3+}$ exists above $T_{V}\,$ due to significant intersite Coulomb
repulsion and frustration on the B-site sublattice.\cite{anderson1956} The
short-ranged electron correlations maintain local charge "neutrality" (2.5+
average valence on each tetrahedron), thereby restricting charge hopping and
conductivity.\cite{ihlelorenz1986} Eventually, Coulomb repulsions win out at
low temperatures, resulting in long-range CO of Fe$^{2+}$ and Fe$^{3+}$
\cite{verwey1947} in a process reminiscent of Wigner
crystallization.\cite{mott1967} Weak elastic and orbital interactions
\cite{leonov2004} induce monoclinic lattice distortions whose complexity
obscures the details of the CO\ state.\cite{zuo1990} Even the validity of the
CO model has been questioned.\cite{garcia2000} However, recent neutron
\cite{wright2002} and resonant x-ray scattering measurements
\cite{nazarenko2006} appear to converge on fractional CO.%
\begin{figure}
[b]
\begin{center}
\includegraphics[
trim=0.5in 0.5in 0.915370in 1.078753in,
scale=0.6
]%
{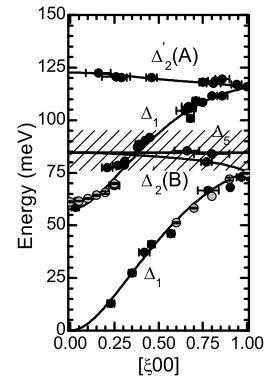}%
\caption{Spin wave dispersion of magnetite above $T_{V}.$ \ Black (gray)
symbols are inelastic neutron scattering data from MAPS (HB-3). \ Lines are
results from a Heisenberg model with parameters discussed in the main text.
The hatched area contains very broad B-site spin waves.}%
\label{fig1}%
\end{center}
\end{figure}

\begin{figure*}[tbp] \centering
{\includegraphics[
]%
{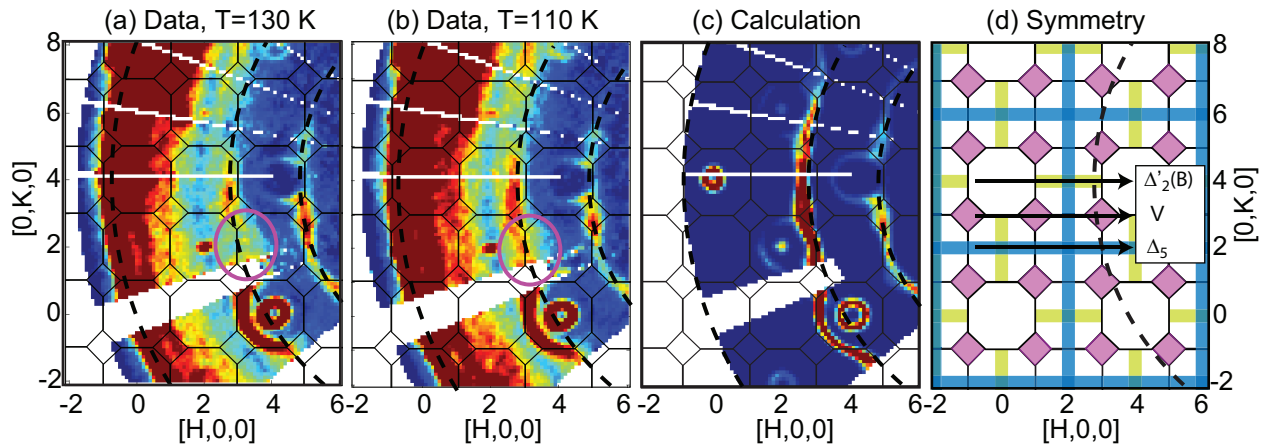}%
}
\caption{(a) Neutron intensity in the $[HK0]$ plane at 130 K.  The color scale indicates neutron
intensity (red the highest). Thin lines show the Brillouin zone boundaries of the cubic phase. Dashed
lines show 0, 80, and 120 meV energy transfers (from left to right). The purple circled region indicates extremely
broad and weak $\Delta_{5}$ symmetry optical spin waves at 80 meV. (b) The 110 K data set below $T_{V}$.
In the purple circled region, the $\Delta_{5}$ mode is more pronounced.  (c) Calculation of the MAPS spin wave
spectra using a Heisenberg model with $J_{BB}$=0.44 meV. Data and calculations have been smoothed by a boxcar averaging procedure.
(d) Symmetry of B-site optical spin waves in different Brillouin zones at 80 meV; blue ($\Delta_{5}$),
green ($\Delta'_{2}$(B)), purple (V). Energy cuts shown in fig. 3 are indicated by arrows.}\label{TableKey}%
\end{figure*}%
In this Letter, we provide strong evidence for Anderson's original picture of
the Verwey transition in magnetite as arising out of short-range electron
correlations in the mixed valent (MV) phase. Valence fluctuations occurring on
the B-sublattice modify the magnetic exchange and affect spin waves
propagating on Fe B-sites. Inelastic neutron scattering measurements of spin
waves reveal that B-site optical spin waves are shifted up in energy and
broadened above $T_{V}$ due to ferromagnetic double exchange (DE).
Ferromagnetic DE arises from real charge transfer processes in MV materials, a
good example being the ferromagnetic metallic state in the
manganites.\cite{tokura2006} For fast electron hopping in the band limit, the
average DE should uniformly affect all B-B pairs. However, our results show
that only spin waves of a particular symmetry are affected by DE, implying the
presence of Zener DE \cite{rosencwaig} arising from slow ($\lesssim80$ meV)
electron hopping and restricted by short-ranged charge correlations. The local
valence fluctuation symmetry can be inferred from the data and mirrors the
eventual long-range symmetry of the CO state.\cite{piekarz2006}
Inelastic neutron scattering measurements were performed on the MAPS
instrument at the ISIS facility at Rutherford-Appleton Laboratory on a
single-crystal of Fe$_{3}$O$_{4}$ weighing $\sim$10 grams. Details of sample
preparation and characterization are given elsewhere.\cite{mcq2006} \ The
sample was mounted with cubic $[HK0]$ as the primary scattering plane and data
were collected with an incident neutron energy of 160 meV and the incident
beam at an angle of 25$^{o}$ from the [110] axis. Time-of-flight neutron
spectra were collected at 110 K ($T$ $<T_{V}$) and 130 K ($T$ $>T_{V}$) and
scattered intensities were histogrammed into energy transfer ($\hbar\omega$)
and momentum transfer ($\hbar$\textbf{Q}) bins. Data were subsequently
analyzed using MSLICE \cite{mslice} and TOBYFIT \cite{tobyfit} computer programs.%
The measured neutron scattering spectra from MAPS (and also the HB--3
spectrometer at the High Flux Isotope Reactor at Oak Ridge National
Laboratory) were used to determine the spin wave dispersion in magnetite. In
the cubic spinel phase above \textit{T}$_{V}$, the primitive rhombohedral unit
cell contains six iron atoms (2A and 4B), leading to six spin wave branches.
Figure \ref{fig1} shows the spin wave dispersion along [100] as determined
from Gaussian fits to spin wave modes. The dispersion is well represented by a
Heisenberg model where antiferromagnetic AB superexchange, via oxygen, is
dominant ($J_{AB}$ $=-4.8$ meV) and is responsible for ferrimagnetism.
Ferromagnetic BB exchange arises from a combination of superexchange, DE, and
direct exchange and is an order-of-magnitude weaker ($J_{BB}=$ 0.69 meV). Weak
antiferromagnetic nearest- and next-nearest-neighbor AA exchange is also
present ($J_{AA}^{(1)}=$ -0.35 meV, $J_{AA}^{(2)}=$ -0.2 meV). The dispersion
calculated from a Heisenberg model with exchange parameters above and spins
$S_{A}=2.5$ and $S_{B}=2.25$ is also shown in fig. \ref{fig1}. Branch
symmetries were identified from the model spin wave eigenvectors and are given
the following labels and descriptions; $\Delta_{1}$ (acoustic and steeply
dispersing optic), $\Delta_{2}^{\prime}(A)$ (optic spin wave on A-sublattice),
$\Delta_{2}^{\prime}(B)$ (optic spin wave on B-sublattice), and $\Delta_{5}$
(doubly-degenerate optic spin waves on B-sublattice). Modes of $\Delta_{5}$
and $\Delta_{2}^{\prime}(B)$ symmetry at $\sim$ 80 meV (in the hatched region
of fig. \ref{fig1}) propagate solely on the B-sublattice and were observed to
be very broad and weak, as discussed below.
Figures 2(a) and (b)\ show images of scattering intensity in the $[HK0]$ plane
at 130 K and 110 K, respectively. The arcs and rings in the images correspond
to the intersection of the neutron measurement surface and the dispersion
surfaces. Figure 2(c) shows a calculation of the MAPS data using the
Heisenberg model described above. Calculations simulate the sampling of spin
waves in \thinspace(\textbf{Q},$\omega)$ space performed by MAPS. Features of
the calculated and measured spectra are in good agreement (as anticipated from
fig. 1) with one notable exception. As indicated in Fig.{}{} 2(a), the
measured B-site optical spin wave branch at $\sim$80 meV is poorly defined in
the region between [220] and [420] at 130 K.\ Comparison with the 110 K
spectrum illustrates two features; (1) the B-site modes sharpen up below
$T_{V}$ in better agreement with the cubic model (despite the lowering of
crystallographic symmetry), (2) other spin wave branches are essentially
unaffected, indicating that $J_{AB}$ and $J_{AA}$ are insensitive to the
Verwey transition.
\begin{figure}
[tb]
\begin{center}
\includegraphics[
trim=0.389539in 0.55in 0.390088in 0.552328in,
scale=0.8
]%
{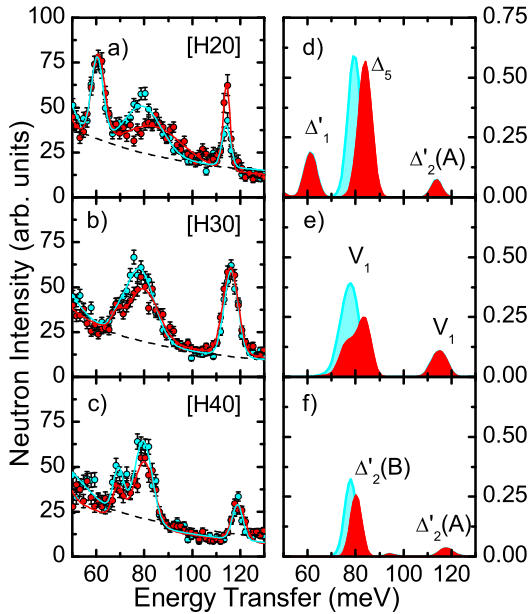}%
\caption{Inelastic neutron scattering intensity as a function of energy
transfer for cuts along;\ (a) $[H20]$, (b) $[H30]$, and (c) $[H40]$.\ Blue
(red)\ symbols indicate measurements done at $T=$ 110 K (130 K). Blue (red)
lines are Gaussian fits to the 110 K (130 K) data. The dashed line is the
background estimated from cuts along $[H60]$. Panels (d) - (f) show the
identical cuts as (a) - (c) as calculated from a Heisenberg model for
magnetite with $J_{BB}=0.44$ meV (blue) and $J_{BB}=0.69$ meV (red).\ Spin
wave modes are labelled by symmetry.}%
\label{fig3}%
\end{center}
\end{figure}
These observations are supported by energy cuts through the MAPS data at
different values of $K$ and different temperatures, as shown in fig. 3. To
improve statistics, cuts are obtained by averaging neutron intensity over a
range of reciprocal space.\ Fig. 3 shows the following cuts made in the range
$-0.25\leq L\leq0.25$; $[H20]~(1.75\leq K\leq2.25)$, $[H30]~(2.75\leq
K\leq3.25),$ and $[H40]~(3.75\leq K\leq4.25)$. These reciprocal space cuts
allow for specific symmetry selection of the B-site optical spin waves, as
shown in fig. 2(d). In the constrained $(\mathbf{Q},\hbar\omega)$ measurement
surface of MAPS, $H$ is a function of $\hbar\omega$, as can be ascertained
from the comparison of figs. 2 and 3. The energy resolution
full-width-at-half-maximum (FWHM) is calculated to be 3-4 meV for
dispersionless modes in the energy range shown (using the TOBYFIT program). In
all cuts, spin wave modes at 60 and 117 meV are independent of temperature and
resolution limited, while $\sim$80 meV modes are broader than the resolution.
The energy cut along the $[H20]$ direction in fig. 3(a) shows that the
$\Delta_{5}$ symmetry branch at $\sim$80 meV shifts up in energy and becomes
severely broadened above $T_{V}$, as also apparent in fig 2. Gaussian fits to
the data indicate that the mode energy shifts from $79.6(5)$ meV to $84(1)$
meV and the FWHM broadens by $60\%$ from 17 to 28 meV above $T_{V}$.\ The
B-site spin waves along $[H30]$ in fig. 3(b) ($\Delta_{2}^{\prime}(B)$
symmetry) and $[H40]$ in fig. 3(c) ($V$ symmetry\cite{vsymmetry}) also have
broad lineshapes with FWHM of 14 meV and 10 meV, respectively, but relatively
little temperature dependence.\ We also note the weak peak near 70 meV in
figs. 3(b) and 3(c)\ is likely to be a phonon, based on examination of
higher-\textbf{Q} data.
The anomalous B-site spin waves can be explained by consideration of
ferromagnetic DE. The conduction electron associated with the Fe$^{2+}$ ion is
forced to remain oppositely aligned to the Fe$^{3+}$ 3d$^{5}$ core spin by the
Pauli exclusion principle (effectively $J_{Hund}\longrightarrow\infty)$. Since
the B-sublattice is already ferromagnetically aligned due to strong $J_{AB}$,
spin polarized conduction results in DE\cite{loos2002} and leads to an
increase in $J_{BB}$. In turn, the energy of B-site optical spin waves
increases. The enhancement of DE above $T_{V}$ can be estimated from the
energy of the dispersionless $\Delta_{5}$ mode, $6J_{AB}S_{B}+8J_{BB}S_{B}$.
Since $J_{AB}$ does not change,\cite{mcq2006} the shift of 4.4 meV implies
that $J_{BB}$ increases from 0.44 meV to 0.69 meV. The energy cuts are
compared to the Heisenberg model structure factor calculation in figs.
3(d)-(f) with $J_{BB}=0.44$ meV and $0.69$ meV. Comparison to calculation
shows that; (1) all B-site spin wave peaks are much weaker and broader than
Heisenberg model predictions, (2) changing $J_{BB}$ in the model shifts the
energy of all mode symmetries, in disagreement with the observation that only
the $\Delta_{5}$ mode has a large temperature dependence. \ Structure factors
in the energy range $75\leq\hbar\omega\leq90$ meV\ and $115\leq\hbar\omega
\leq120$ meV and $L$-range $-0.25\leq L\leq0.25$ are shown in fig. 4 as a
function of $K$.\ Fig. 4(a) further illustrates the anomalous $\Delta_{5}$
mode, which shows a minimum in the measured structure factor whereas the
calculation predicts a maximum.\ Fig. 4(b) indicates that A-site optical spin
waves have little temperature dependence and agree quite well with Heisenberg calculations.%
\begin{figure}
[pt]
\begin{center}
\includegraphics[
trim=0.5in 0.5in 0.75in 0.75in,
scale=0.9
]%
{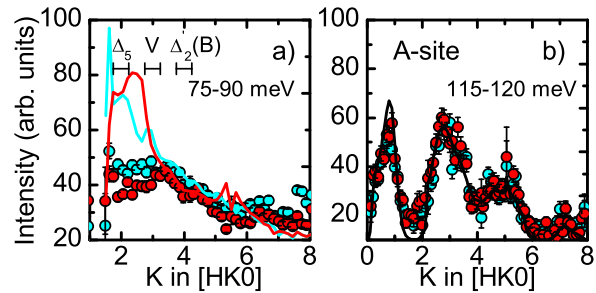}%
\caption{(a) Neutron structure factor for B-site optical spin waves summed
from 75-90 meV. \ Red (blue) circles are data at 130 K (110 K). \ Red (blue)
lines are calculated from a Heisenberg model with $J_{BB}=$0.69 (0.44)\ meV.
\ Horizontal bars indicate dominant symmetry. \ (b) Structure factor for
A-site optical spin waves from 115 -120 meV. \ Black line is the result of
Heisenberg model.}%
\label{fig4}%
\end{center}
\end{figure}
It is clear that stiffening and extreme broadening of $\Delta_{5}$ modes
above $T_{V}$ cannot be explained by uniformly enhanced BB coupling as might
be expected from a band approach. The failure of a band model is reiterated by
estimating the DE arising from metallic conductivity, $J_{DE}\sim nt/2\sqrt
{2}S_{B}^{2}$ \cite{millis1995,furukawa1996} where $n\approx1/2$ is the
concentration of charge carriers in the conduction band and $t$ the electron
hopping integral. The effective one-band hopping integral from LSDA
calculation is $\sim130$ meV\cite{zhang1991}, leading to an estimate of
$J_{DE}\sim5$ meV that is significantly larger than the observed value
($J_{DE}\approx J_{BB}(T>T_{V})-J_{BB}(T<T_{V})=0.25$ meV). This disagreement
emphasizes the importance of charge correlation and/or polaronic effects above
$T_{V}$ that lead to activated conduction and a reduction of the effective
hopping integral. This picture is supported by optical conductivity
measurements showing that the $\sim$100 meV electronic energy gap below
$T_{V}$ (which suppresses DE) is replaced by a pseudogap above $T_{V}%
$.\cite{degiorgi1987,park1998}\ In this limit, the Zener DE picture of slow
and local charge hopping between distinct valence states above $T_{V}$ is
appropriate and the closing of the electronic gap allows the coupling of spin
waves to valence fluctuations leading to broadened spinwave lineshapes.\ The
predominant $\Delta_{5}$ spin wave symmetry would then imply that valence
fluctuations have the same symmetry and occur along the quasi-one-dimensional
[110] B-site chains, as illustrated in fig. 5. The spin wave changes the
alignment of neighboring spins, thereby impeding valence fluctuations which
favor parallel spins.
The presence of slow $\Delta_{5}$ valence fluctuations coupling to the lattice
has also been observed with neutron diffuse scattering and occur well above
$T_{V}$ \cite{shapiro1976}. In the MV phase, these local symmetry-breaking
charge correlations are short-ranged. While we interpret our data based on the
symmetry assignments for modes along the [100] direction, broadening
everywhere in the Brillouin zone indicates that the coupled B-site charge and
spin fluctuations are local. The resulting picture of valence fluctuations
constrained by short-range ordering of Fe valences is similar to that proposed
originally by Anderson.\cite{anderson1956} \ In addition, the implication of
slow electron hopping ($\lesssim$80 meV) supports more detailed theories of
magnetite that invoke polaronic behavior for the Fe$^{2+}$ charge
carriers.\cite{ihlelorenz1986} Spin-charge coupling is large and may
contribute to the polaronic binding energy.\ Finally, the $\Delta_{5}$
symmetry of charge correlations are similar to the eventual frozen pattern of
the CO state, which has been argued to be a combination of $\Delta_{5}\oplus
X_{3}$ cubic representations.\cite{piekarz2006} \ \ \ %
\begin{figure}
[ptb]
\begin{center}
\includegraphics[
scale=0.8
]%
{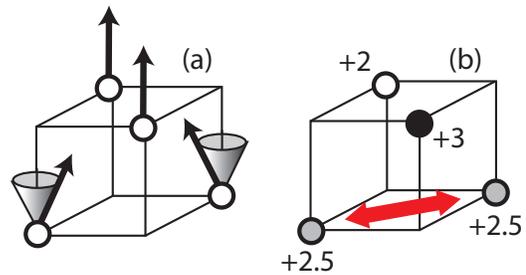}%
\caption{Schematic drawing of (a) spin and (b) charge modes on the
B-sublattice with $\Delta_{5}$ symmetry. \ In (b), the heavy arrow indicates
direction of charge fluctuations.}%
\label{fig5}%
\end{center}
\end{figure}
\appendix{}
\begin{center}
ACKNOWLEDGMENTS
\end{center}
RJM would like to thank S. Satpathy for useful discussions. \ Work is
supported by the U. S. Department of Energy Office of Science under the
following contracts; Ames Laboratory under Contract No. DE-AC02-07CH11358, Oak
Ridge National Laboratory, which is managed by UT-Batelle LLC, under Contract
No. DE-AC00OR22725.
\bibliographystyle{apsrev}
\bibliography{acompat,mcq}

\begin{thebibliography}{26}
\bibitem[{vsy()}]{vsymmetry}
\bibinfo{note}{{\sc V} symmetry direction is $[1\xi0]$, on the Brillouin zone
  face}.
\end{thebibliography}
\end{document}